\documentclass{aa}
\usepackage{graphicx}
\usepackage[table]{xcolor}
\usepackage[dvipsnames]{xcolor}
\usepackage[caption=false]{subfig}
\usepackage{txfonts}
\usepackage{multirow}
\usepackage{multicol}
\usepackage{tabu}
\usepackage{textgreek}
\usepackage{float}
\usepackage{natbib}
\usepackage{enumitem}
\usepackage[pdfauthor={},pdftitle={},colorlinks=true]{hyperref}

\newcommand{\heim}{\textsc{heimdall}}
\newcommand{\dm}{\,pc\,cm$^{-3}$}
\newcommand{\rfibye}{{\tt RFIbye.py}}
\newcommand{\psrcat}{\mbox{\textsc{psrcat}}}
\newcommand{\hms}[3]{\ensuremath{{#1}^\mathrm{h}{#2}^\mathrm{m}{#3}^\mathrm{s}}}
\newcommand{\dms}[3]{\ensuremath{{#1}^\mathrm{\circ}{#2}\mathrm{'}{#3}\mathrm{''}}}

\begin{document} 

   \title{The Northern High Time Resolution Universe pulsar survey}

   \subtitle{III. Single-pulse search continuation, follow-up observations, and initial results}

   \author{L.~J.~M. Houben\inst{1}\fnmsep\inst{2}\fnmsep\thanks{\email{l.houben@astro.ru.nl}}
           \and H. Falcke\inst{1}\fnmsep\inst{3}\fnmsep\inst{2}
           \and L.~G. Spitler\inst{2}
           \and \\ E.~D. Barr\inst{2}
           \and M. Berezina\inst{2}
           \and D.~J. Champion\inst{2}
           \and R. Karuppusamy\inst{2}
           \and M. Kramer\inst{2}\fnmsep\inst{4}}

   \institute{Department of Astrophysics/IMAPP, Radboud University, PO Box 9010,
              6500\,GL Nijmegen, The Netherlands
        \and
              Max-Planck-Institut f\"ur Radioastronomie, Auf dem H\"ugel 69,
              D-53121 Bonn, Germany
        \and
              ASTRON, the Netherlands Institute for Radio Astronomy, Oude Hoogevensedijk 4,
              7991 PD Dwingeloo, The Netherlands
        \and
              Jodrell Bank Centre for Astrophysics, University of Manchester,
              Alan Turing Building, Oxford Road, Manchester M13 9PL, UK}

   \date{}

   \abstract{We continued the search for single pulses (SPs) in the northern part of the all-sky High Time Resolution Universe survey, whose aim is to detect pulsars and other radio transients. This search is now about 21\% complete and has yielded the first discovery of a fast radio burst (FRB) with the 100\,m Effelsberg Radio Telescope. FRB\,20110220A was detected with an S/N-optimised dispersion measure of 501.0\dm\ and a width of 11.9 $\pm$ 3.5\,ms, for a fluence of 0.6 $\pm$ 0.1\,Jy\,ms. We obtained the first L-band detection of the rotating radio transient (RRAT) J2028+28, from which we obtained upper limits on the source's period and burst rate, as well as an improved position. We also discovered a new RRAT, J0404+53, which had previously been reported as an isolated SP candidate. Eight new SP trains and 272 faint isolated SP candidates were detected too. We used these candidates to demonstrate that their all-sky detection rates depend on Galactic latitude and longitude. This direction dependence suggests the existence of a faint Galactic SP population.}


   \maketitle

\section{Introduction}
The discovery of the first periodically emitting pulsar \citep{hbp+1968} triggered considerable interest in transient radio astronomy and marked the birth of this new field within astronomy. Subsequent research broadened the range of sources studied. Pulsars were discovered from which no radio pulses were observed for several rotational periods in a row, the so-called nulling pulsars \citep{bac1970}, or from which pulses were even more irregularly observed: rotating radio transients (RRATs; \citealt{mll+2006}). Later, new impulsive radio sources were discovered through single-pulse (SP) searches of archival radio data. Fast radio bursts (FRBs; \citealt{lbm+2007}) were discovered by searching the time domain of radio data for SPs, and similar searches in the image domain revealed the existence of long-period transients (\citealt{hzb+2022}). Searches for SPs in archival radio data might even continue to unravel new phenomena, as discussed later on in this paper.

\citet{hfs+2026} created a pipeline to search the archival data of the northern part of the High Time Resolution Universe (HTRU) survey for SPs. This survey \citep{bck+2013}, conducted with the 100\,m Effelsberg Radio Telescope, complements the southern part of the HTRU survey \citep{kjv+2010}, for which the 64\,m Parkes Radio Telescope was used. Together, the two surveys intend to map the entire radio sky in high time and frequency resolution with the aim of discovering new and impulsive radio sources.

Since pulsars are mainly located in or near the Galactic plane, observations pointed in these directions are expected to have the highest chance of capturing new pulsar systems. The HTRU-North survey has therefore focused its observation campaigns on the Galactic latitudes \mbox{|$b$| < 15$^{\circ}$} with 180\,s integration times. By periodicity-searching half of the obtained data, 19 new pulsars were discovered \citep{bck+2013,ber2019}. However, the event rates of nulling pulsars and RRATs might be too low within the 180\,s long observations to render such sources detectable with periodicity searches. Not being detectable via periodicity searches is in fact a defining characteristic of RRATs. Since detectability is highly dependent on a source's rotational period and the length of an observation, the RRAT label might more represent a detection limitation rather than a physical phenomenon as described by \citet{dcm+2009} and \citet{bbj+2011}.

The newly created SP search pipeline increases the chance of detecting such sources within the HTRU-North data and enables the detection of fast transients such as FRBs. This pipeline has already been used to analyse the data of an initial $\sim$1500 sky pointings (PTs). The continuation of the SP analysis on the available HTRU-North data is reported in this paper. The current progress of the SP search is described in Sect. \ref{sec:status}. In response to the results presented in \citet[hereafter HFS26]{hfs+2026}, follow-up observations were performed to investigate the origin of the reported SP trains and candidates. Section \ref{sec:follow-up} elaborates on how these observations were performed. The outcomes from the continued SP search and follow-up observations are presented in Sect. \ref{sec:results}, and their implications are discussed in Sect. \ref{sec:discussion}. The paper is summarised in Sect. \ref{sec:conclusions}.
\section{Search status} \label{sec:status}
For the HTRU survey, the sky is divided into three regions based on Galactic latitude: the low-latitude (low-lat) region with \mbox{|$b$| < 3.5$^{\circ}$}, the mid-latitude (mid-lat) region with \mbox{|$b$| < 15$^{\circ}$}, and the high-latitude region with \mbox{|$b$| > 15$^{\circ}$}. The observation time of a PT in a particular region is chosen such that the northern and southern surveys have similar sensitivity limits. For the HTRU-North observations, the integration times are 1500, 180, and 90\,s, respectively. Currently, only 57 low-lat PTs with an integration time of 1500\,s exist. Due to the mid-lat observations covering the low-lat region as well, practically all available PTs in the low-lat region have an integration length of 180\,s. For clarity, mentions of mid-lat PTs shall, from now on, refer to PTs with a latitude of \mbox{3.5$^{\circ}$ < |$b$| < 15$^{\circ}$}, and PTs with a latitude \mbox{|$b$| < 3.5$^{\circ}$} are referred to as low-lat PTs regardless of their observation length.

Using Effelsberg's 21\,cm multibeam receiver and its polyphase filterbank backend, the HTRU-North observations resulted in \textsc{sigproc}\footnote{\url{https://sigproc.sourceforge.net/}} filterbank files, one for each of the receiver's seven beams, with a time resolution of 54.61\,\textmu s and a recorded bandwidth of 300\,MHz centred around 1360\,MHz split into 512 channels. The obtained filterbank files were searched for pulsars, as described by \citet{bck+2013}, and SPs with the SP search pipeline discussed in HFS26. The functionality of this pipeline was characterised through the injection of fake SPs into the beam files of an initial $\sim$1500 PTs. HFS26 presented the results of the analysis of these initial PTs and the conclusions drawn from the resulting characterisation.

Since then, the HTRU-North SP search pipeline was set to its production mode. In this mode, HTRU-North filterbank files were first retrieved from the archive, where they are stored, before being analysed with the SP search algorithm \heim\footnote{\url{https://sourceforge.net/projects/heimdall-astro/}} \citep{bbbf2012}. Each file was searched for SPs with \heim\ twice, once before and once after the mitigation of radio frequency interference (RFI). In this way, information is obtained about the degree of RFI contamination of the data and the reliability of the SP candidates accordingly. Only the candidates with a member count of three or higher, i.e. the number of boxcars and/or dispersion measure (DM) trials that \heim\ grouped into a single candidate, were further taken into consideration. In contrast to the candidates found in the initially analysis PTs, candidates detected with the pipeline set to its production mode were no longer classified by the deep-learning classifier \textsc{fetch} \citep{aab+2020}. Its classification with pre-trained models proved not to outweigh the time and resource costs to get a classification score for the few candidates that remained after RFI mitigation. In production mode, no fake SPs are injected. In the following, PTs processed with the pipeline in its production mode are referred to as the `production PTs'.

\begin{figure*}[h!]
\centering
\includegraphics[width=\textwidth]{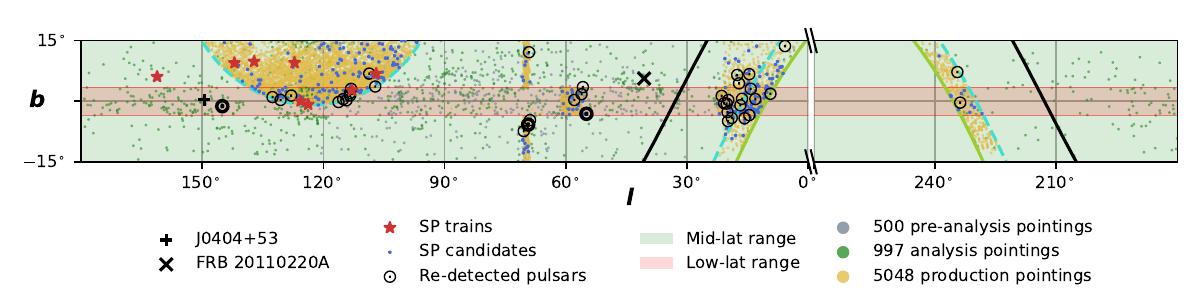}
   \caption{All currently processed HTRU-North PTs at the Galactic positions they cover. A PT's relative area on the plot is equal to the area of the Effelsberg 7-beam receiver's beam pattern on the sky. Overlaid are the new discoveries, SP trains, SP candidates, and known pulsar and RRAT re-detections. The pulsars shown in boldface are, from left to right, B0329+54, J2028+28, and B1942+17. The solid lime and dashed cyan lines are the declination limits of Effelsberg and FAST, respectively.}
   \label{fig:pointing_skymap}
\end{figure*}

The HTRU-North survey shares sky coverage with other radio transient searching surveys \citep{sha+2011,sta2018,sal+2019,lbb+2020,hww+2021,myw+2022,dcm+2023,vko+2023}. For the continuation of the HTRU-North SP search, PTs were prioritised that are not visible with the Five-hundred-meter Aperture Spherical Telescope (FAST; \citealt{nlj+2011}). This was done to increase the chance of finding yet undiscovered sources, as FAST is more sensitive than Effelsberg and thus more capable of detecting faint sources, as shown by its recent discovery of 76 RRATs~\citep{zhx+2023}. Most production PTs therefore lie above or below the magenta line in Fig. \ref{fig:pointing_skymap}, which indicates FAST's declination limit. The 600 production PTs around a Galactic longitude of 70$^{\circ}$ are additionally analysed in order to deduce a possible longitude dependence in the detection rate of isolated SP candidates (see Sect. \ref{sec:sp-cands}).

With the SP search pipeline, 5048 production PTs were analysed in addition to the 1497 previously analysed PTs. This brought the total amount of processed PTs to 2357\,hours\footnote{This value includes 18 true low-lat PTs and the fact that some PTs have been observed multiple times, raising their total observation time.}, or $\sim$21\% of the available HTRU-North data. The PTs' Galactic distribution is given in Fig. \ref{fig:pointing_skymap}.

\section{Follow-up observations} \label{sec:follow-up}
From the 1497 previously analysed PTs, several SP trains and isolated SP candidates of unknown origin were obtained. To better interpret these candidates, follow-up observations were performed in an attempt to re-detect them. These observations were done with the 21\,cm receiver of Effelsberg. Since the last HTRU-North observation, Effelsberg has been updated with the Effelsberg Direct Digitisation (EDD) system \citep{bwb+2023}. The EDD system consists of a modular frontend, which digitises an analogue signal directly at the receiver, and a backend that performs the signal processing in software. The backend can handle several observing modes, which would previously have required several single-purpose backends. In pulsar search mode, the EDD replaces the Effelsberg Pulsar Fast Fourier Transform Spectrometer (PFFTS) backend that was used for the HTRU-North observations. Therefore, the follow-up observations' data differ slightly from the HTRU-North data.

For the follow-up observations, the 21\,cm receiver's two polarisation channels were still summed to obtain total intensity data. However, these data now have a time and frequency resolution of 51.2\,\textmu s and 0.78\,MHz, respectively. The frequency resolution arises from splitting the 400\,MHz digitally sampled bandwidth, centred around 1.4\,GHz, in 512 frequency channels. Only 300\,MHz of this bandwidth was usable, because the receiver's bandwidth remained the same. To prevent intra-channel dispersion smearing in the wider frequency channels, we coherently de-dispersed the follow-up data with the DM of the observed candidate. Data were stored in the Flexible Image Transport System (\textsc{fits}) format \citep{pcp+2010} and later converted to the filterbank format to allow the application of all the previously developed tools to these data.

Applying \rfibye\footnote{\url{https://gitlab.com/houben.ljm/rfibye}} \citep{hfs+2026} to the data was needed to excise RFI to a level that it was clean enough to be searched for SPs. Especially because the RFI environment at the telescope has changed in the ten years since the HTRU-North data were recorded. On average, $\sim$5\% of data samples of an HTRU-North PT's filterbank file must be removed or replaced with random noise. For the follow-up observations, 60 -- 80\% of data samples are considered bad by \rfibye. In particular, frequency channels above 1440\,MHz are affected by the new RFI environment and have become unusable for the 21\,cm receiver. This reduces the effective bandwidth of the follow-up data to $\sim$180\,MHz in most cases, which is still hampered by a large variety of impulsive emission that is difficult to remove. Tens to hundreds of false positive events were therefore detected after RFI mitigation (see e.g. the re-detection plot of B1942+17 in Appendix \ref{sec:A.figs&tabs}). As a result, the new follow-up observations are less sensitive than the original HTRU-North data, making it harder to identify true positive events.

Given the uncertainty in the candidates' positions, the position of a candidate's detection beam was re-observed with four consecutive PTs of the 21\,cm receiver's centre beam. The first of these PTs was centred on the sky location of the initial detection beam. The three following PTs were then centred on the full width half maximum of the first PT, such that their centres formed an equilateral triangle (see Fig. \ref{fig:up_pos}). In addition to a possible boost in S/N due to the applied gridding, this also enabled a better determination of a candidate's position when it is re-detected with one of the outer PTs.

\begin{table}
\small
\caption{Schedule of the follow-up observations performed in 2025.}
\label{tab:obs_schedule}
\centering
\begin{tabular}{l | c c c c c c c c}
    C016 & & & & 7.5 & & & & \\
    T054.9-03.0 & & & & \textbf{15} & & & & \\
    T055.4+00.2 & & & & & & 10 & & \\
    T059.5+00.8 & & & & 10 & & 10 & & \\
    T052.3+07.7 & & & & 15 & & 13.5 & & \\
    T121.8-03.5 & 7.5 & 10 & & & & 13.5 & & \\
    T093.1-08.5 & & & & & 15 & & 10 & \\
    T065.1-10.7 & 7.5 & & 10 & & & & 10 & \\
    J2028+28 & & & & & 10 & & \textbf{15} & \\
    T102.1+01.4 & & & & & & & 15 & \\
    C116 & & & & & & & & 20 \\
    T142.0+9.46 & & & & & & & & 15 \\
    \noalign{\smallskip}
    \hline
    \noalign{\smallskip}
     & \rotatebox[origin=l]{80}{{\tiny 04 Apr}} & \rotatebox[origin=l]{80}{{\tiny 23 Jul}} & \rotatebox[origin=l]{80}{{\tiny 24 Jul}} & \rotatebox[origin=l]{80}{{\tiny 19 Aug}} & \rotatebox[origin=l]{80}{{\tiny 20 Aug}} & \rotatebox[origin=l]{80}{{\tiny 27 Aug}} & \rotatebox[origin=l]{80}{{\tiny 28 Aug}} & \rotatebox[origin=l]{80}{{\tiny 06 Oct}}\\
\end{tabular}
\tablefoot{Indicated times are for a single scan, given in minutes. Bold-faced times indicate when SPs of the observed candidate were re-detected. These candidates are taken from HFS26 and Table \ref{tab:sp_trains}.}
\end{table}

Details of the performed follow-up observations are given in Table \ref{tab:obs_schedule}. An effort was made to observe the candidates on at least two different days, separated in time by a week to account for temporal changes in the source's flux density or the level of RFI. In the preparation of the follow-up observations, observations were scheduled such that targets were not observed in directions identified as often hampered by RFI (see HFS26). Nevertheless, the data from the follow-up observations are still heavily contaminated with RFI. For the new RFI environment at the telescope's site, no time or celestial sky location dependence could be determined. Specific types of RFI are picked up at random moments throughout the observations.

\section{Results} \label{sec:results}
In this section, results obtained from the continued HTRU-North observations are interleaved with results from the follow-up observations. To prevent confusion about which results have been obtained from which effort, we use the term `PT' only to refer to the HTRU-North PTs. The follow-up PTs are called `scans'. Since just the central beam of Effelsberg's 21\,cm receiver was used for the performed scans, any mention of a beam refers to a beam of an HTRU-North PT. A follow-up observation is thus the collection of four scans, where scan 0 indicates the centre scan of the performed gridding and scans 1 to 3 the outer scans numbered from the most northerly scan in a clockwise direction.

\subsection{Re-detections} \label{sec:re-detections}
In addition to the known pulsars and RRAT re-detected in HFS26, 37 additional known pulsars and 3 known RRATs have been re-detected through their SPs in the production PTs (see Table \ref{tab:re-detections}). In total, 62 known neutron stars have now been re-detected in the HTRU-North PTs processed so far.

The data of these known pulsars were folded with the DM and period of the pulsars as listed in the  Australia Telescope National Facility Pulsar Catalogue (\psrcat\footnote{\url{https://www.atnf.csiro.au/research/pulsar/psrcat}}; \citealt{mhth2005}), and 28 were re-detected with an integrated pulse profile with an S/N > 6.5. As expected, no integrated pulse profile was detected for the three RRATs. Twelve sources would thus have been missed if the data were only periodicity searched for pulsars, which endorses the use of SP searches alongside periodicity searches.

The 12 known pulsars with an integrated S/N below 6.5 were missed due to several reasons. They were detected either on the edge of or outside their detection beam's full width at half maximum, because of which only their brightest pulses were detected \citep{dcm+2009}. The pulsars are known to null \citep{bjb+2012}, or their period in combination with their mean flux density at 1.4\,GHz and off-axis detection renders their average emission undetectable. The detected SPs could nevertheless be confirmed to originate from the listed known pulsars based on the Galactic position of their detection beam, the DM of the SPs, and their arrival times being integer multiples of the pulsar's period.\newline

\begin{table}
\caption{Additional known pulsar and RRAT re-detections within the HTRU-North production PTs.}
\label{tab:re-detections}
\centering
\begin{tabular}{l l l | c}
    \hline
    \noalign{\smallskip}
    & Pulsars & & RRATs \\
    \noalign{\smallskip}
    \hline
    \noalign{\smallskip}
     J0137+6349 & B1822-09 & B1942+17$^{\dagger}$ & \\
     B0154+61 & B1822-14 & B2020+28 & \\
     J0215+6218 & B1823-11 & J2036+2835 & \\
     B0329+54$^{\dagger}$ & B1826-17 & B2224+65 & \\
     B0727-18 & B1828-11 & B2227+61 & \\
     B0756-15 & J1830-1135 & J2319+6411 & J1753-12 \\
     B1706-16 & J1834-1710 & B2323+63 & J1840-1419 \\
     J1759-1029 & J1837-1243 & J2326+6243 & J2028+28$^{\dagger}$ \\
     J1759-1956 & J1845-1351 & B2334+61 & \\
     B1804-12 & J1913+3732 & J2343+6221 & \\
     J1805-1504 & B1919+21 & B2351+61 & \\
     B1809-173 & J1929+2121 & & \\
     B1815-14 & J1938+2213 & & \\
    \noalign{\smallskip}
    \hline
\end{tabular}
\tablefoot{$^{\dagger}$ See the sub-subsections in Sect. \ref{sec:results} for details.}
\end{table}

The here reported re-detection of J2028+28 is the first time this RRAT is observed at L-band. Lower frequency detections are listed on the Canadian Hydrogen Intensity Mapping Experiment (CHIME) FRB Galactic sources web page\footnote{\url{https://www.chime-frb.ca/galactic}}, where three bursts are reported at a consistent position and DM on three separate days. Because the detections were made with different observations, they could not determine a period for J2028+28. However, in a single beam of an HTRU-North PT, two SPs were detected from J2028+28 separated by 1.764\,s, which places an upper limit on its period. Folding the data with this period did not reveal any average emission. It thus seems to show RRAT behaviour, only emitting SPs infrequently.

\begin{figure}
\centering
\includegraphics[width=0.48\textwidth]{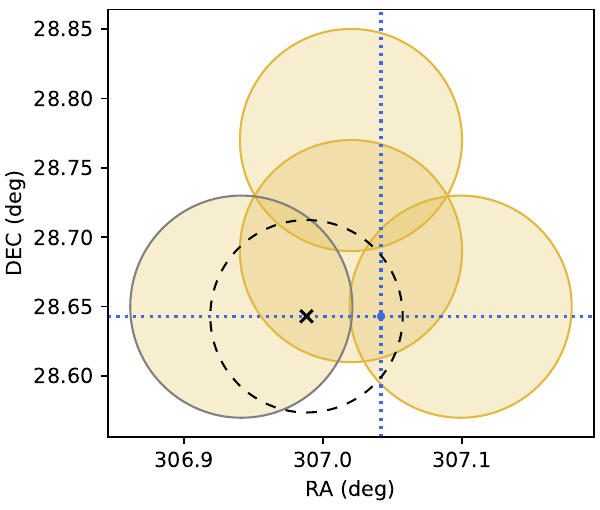}
   \caption{Grid pattern used to re-detect SPs from J2028+28 within its HTRU-North detection beam (the centre yellow circle). One SP was observed in the grey encircled scan. The dotted blue lines represent the semi-major and minor axes of CHIME/FRB's tied-array beam, which is larger than the plotted area, centred on their reported position of J2028+28. The black cross marks the updated position for J2028+28 with an error radius of 0.07$^\mathrm{\circ}$. Note that the lower scans seem to be spaced farther apart compared to the uppermost scan. This is a projection effect and is incorporated in the given error radius.}
   \label{fig:up_pos}
\end{figure}

Because of the sparse information available, J2028+28 was included as one of the follow-up targets (Table \ref{tab:obs_schedule}). The position of the detection beam was re-observed on August 20 and 28, 2025, with scans of 10 and 15 minutes, respectively. A single SP with an S/N of 9.43 was detected in the last scan. This constrains the position of J2028+28 to RA = \hms{20}{27}{57}, Dec = \dms{28}{38}{35} with an error radius of 0.07$^\mathrm{\circ}$ (see Fig. \ref{fig:up_pos}). For this updated position, the sky overlap of the detection beam and scan was considered, together with the reported position of J2028+28 by the CHIME/FRB collaboration.

After de-dispersing the three bursts to optimise their S/Ns, an average DM and burst width were found of \mbox{127 $\pm$ 2\,\dm} and \mbox{6.1 $\pm$ 0.7\,ms}, respectively. Since the Galactic position of J2028+28 is covered with two scans per follow-up observation, a total of 53 minutes were spent observing this source (including the observation that led to its discovery). With three SPs detected in this time, its burst rate is roughly 0.3 bursts~h$^{-1}$.

\subsection{Candidate pulse trains} \label{sec:sp-trains}
HFS26 defines candidate SP trains as events that comply with the following criteria:
\begin{enumerate}
    \item They consist of a minimum of three SPs.
    \item They are detected in a single beam of an HTRU-North PT.
    \item The individual SPs' DMs deviate from the SP train's average DM by less than 1.5 times its DM standard deviation.
    \item The individual SPs follow an underlying period.
\end{enumerate}
Under these conditions, eight SP trains were found in the initially analysed PTs. However, the S/N values of their individual SPs, between 6.5 and 8, made it hard to interpret their significance. Therefore, the SP trains were reported as candidate SP trains and were re-observed at least once from April to October 2025 (see Table \ref{tab:obs_schedule}). To see if the SP trains were re-detected, the follow-up observations' data were searched with the SP pipeline.

Nine SPs were detected in the second scan of the re-observation of T054.9-03.0 on August 19, 2025. Their DMs matched that of pulsar B1942+17 and folding the data between 50 and 350\,s into this scan (Fig. \ref{fig:det_plot_B1942+17}), closest to the position of B1942+17, confirmed them to originate from this pulsar through the detection of an integrated pulse profile. Three more SPs were manually identified in scan 0, though they were missed by \heim\ due to their lower S/Ns.

\citet{lcx2002} reports B1942+17 to be a strong nulling pulsar with a nulling fraction of at least 60\% at 430 MHz. Visual inspection of the follow-up data revealed 28 SPs between the first and seventh SP found by \heim, yielding a nulling fraction of $\sim$80\%. This large nulling fraction, together with the pulsar's long period ($\sim$2\,s) and off-axis HTRU-North detection, explains why only three bursts were initially detected and no average emission could be found to link them to this pulsar. Although T054.9-03.0 does not represent a new pulsar, its detection does show that at least a fraction of the detected faint SPs are true positive events (see Sect. \ref{sec:sp-cands}).

No SPs of the other SP trains were re-detected, though average emission might be detectable due to the longer integration lengths of the follow-up observations compared to those of the HTRU-North observations. The cleaned follow-up data were therefore folded with the periods and averaged DMs measured from the SP trains' individual SPs. In cases where the recording of a scan is affected by periods of strong RFI, the least affected time span is isolated and used for the folding, as was done for the scan of T054.9-03.0. No integrated pulse profiles were found for the other re-observed SP trains either.\newline

The detection of J2028+28 and several known pulsars suggests that the above constraints, under which SPs are considered to belong to a SP train, might be too strict. The re-detection of B0756-15 in particular has called these constraints into question. It was clearly detected in beam 6, but in the centre beam of that PT, only two SPs of this pulsar were detected with an S/N < 8 and a similar DM. Looking solely at the bursts in the centre beam, these look like the initial detected bursts from B1942+17 and J2028+28. What these bursts have in common is a relatively high \heim\ member count, i.e. the number of boxcars and/or DM trials that \heim\ grouped into a single candidate. Although this member count depends on a SP's DM and width\footnote{For specific settings of \heim. If the DM trials are performed with increasingly bigger step sizes, a high DM SP can get detected with fewer DM trials, ergo member count, than a lower DM SP with an equal S/N. The same holds true for the widths of search boxcars used. Narrower bursts can be found with fewer trial boxcar widths than wider bursts.}, along with its S/N and the RFI environment of the data in which it is detected, this metric can be used as an indication of its authenticity. As learned from the injection tests in HFS26, a member count > 10 seems to indicate that a candidate is most likely a true positive event for the HTRU-North SP pipeline. All injections of FRB Morphology I \citep{pgk+2021} with an S/N > 7.5 were detected with a member count > 10. Other candidates with an S/N > 7.5 could either easily be visually identified as RFI or had a member count < 10. Also, pulsar J1759-1029 could be identified based on three detected SPs with a member count of 11, 11, and 12, which supports the notion that this member count threshold is indicative of true positive events.

The above leads to a refined definition of what is considered a SP train. Requirement 2 from HFS26 is relaxed to include SPs detected in adjacent beams. In addition, a SP train can now also consist of two SPs if they comply with these requirements:
\begin{enumerate}
    \item They are detected in a single beam of an HTRU-North PT.
    \item Their DM deviates by no more than 5\,\dm\ from their mean.
    \item Each burst has a member count greater than or equal to five (the same condition raised by HFS26 to consider a detection as an isolated SP candidate).
    \item They are detected in a PT that can be considered free from RFI after RFI mitigation.
\end{enumerate}
With these updated conditions, seven SP trains were identified in the production PTs and one in the initially analysed PTs. The new candidate SP trains have an average DM that places them well within the Galaxy when the Galactic DM contribution is estimated with the NE2001 model \citep{cl2002} for the Galactic locations of the trains. Consulting the Pulsar Survey Scraper (PSS; \citealt{kap2022}), no known source is found within a 0.25$^\mathrm{\circ}$ radius of these locations (the beam separation of the 21\,cm multibeam receiver) with a DM within $\pm$ 25\,\dm\ of the SP trains (the DM difference between T054.9-03.0 and B1942+17). Their locations are indicated in Fig. \ref{fig:pointing_skymap} and their details are listed in Table \ref{tab:sp_trains}.

\subsection{SP candidates} \label{sec:sp-cands}
Along with the detection of additional known pulsars and candidate SP trains, many isolated SPs were detected in the production PTs as well. Applying the same constraints from HFS26, the chance of them originating from instrumental noise or RFI was reduced. Therefore, only SPs with a member count of five or higher, and which were detected in a PT cleared of RFI with \rfibye\ were considered. The resulting 272 isolated SP candidates were cross-referenced with the PSS to determine if they could originate from a known source. For six isolated SP candidates, a known source was found within a 0.25$^\mathrm{\circ}$ radius of a SP candidate's position and within a DM range of $\pm$ 25\,\dm\ of its DM. Except for one burst of B0329+54 (see Sect.~\ref{sec:sp_disc}), all re-detected known neutron stars were observed within this 0.25$^\mathrm{\circ}$ radius, including side-lobe detections that could explain the low S/Ns of the detected isolated SP candidates. Therefore, the other candidates potentially originate from yet-unknown sources. Table \ref{tab:sp_cands} lists all the detected isolated SP candidates, continuing the numbering used in HFS26, and their observed positions are plotted in Fig. \ref{fig:pointing_skymap}.

In HFS26, a possible latitude dependence was observed on the detection rate of isolated SP candidates. More SP candidates were detected in the low-lat PTs than in the mid-lat PTs, even though the amount of analysed data from these regions was the same. Since the low-lat PTs cover the Galactic plane, this latitude dependence could suggest the existence of faint background SP emission from a possible Galactic origin.

\begin{table}
  \small
  \caption{Number of detected Galactic isolated SP candidates in the observed Galactic latitude and longitude regions, and the detection rates deduced from them.}
  \label{tab:sp_rates}
  \centering
  \begin{tabu}{l | c c c c | r}
    \rowfont{\tiny}
    &$l$000-030&$l$030-090&$l$090-150&$l$210-270& \\
    \noalign{\smallskip}
    \hline
    \noalign{\smallskip}
    low-lat & 50 & 20 & 45 & 9 & 11.5$^{+1.8}_{-1.6}$ \\
    mid-lat & 44 & 18 & 85 & 1 & 6.0$^{+0.9}_{-0.8}$ \\
    \noalign{\smallskip}
    \hline
    \noalign{\smallskip}
    & 14.0$^{+2.6}_{-2.3}$ & 9.6$^{+3.0}_{-2.4}$ & 5.9$^{+0.9}_{-0.8}$ & 3.9$^{+2.7}_{-1.8}$ & \\
  \end{tabu}
  \tablefoot{Rates are given in $\times 10^{6}$\,sky$^{-1}$\,day$^{-1}$, with 90\% Poisson errors.}
\end{table}

To determine whether such a Galactic relation exists, the isolated SPs from Table \ref{tab:sp_cands} were grouped into six Galactic longitude and latitude regions. The number of detected candidates in a region was then normalised by the amount of time processed of this region and the field of view of a beam. This yielded the detection rates listed in Table \ref{tab:sp_rates} given in units of sky$^{-1}$\,day$^{-1}$. The rates are reported with their corresponding 90\% Poisson error values \citep{geh1986}. As can be seen from Table \ref{tab:sp_rates}, the SP detection rate is almost twice as high in the low-lat region as in the mid-lat region, confirming the latitude dependence observed in HFS26.

\begin{figure}
\centering
\includegraphics[width=0.48\textwidth]{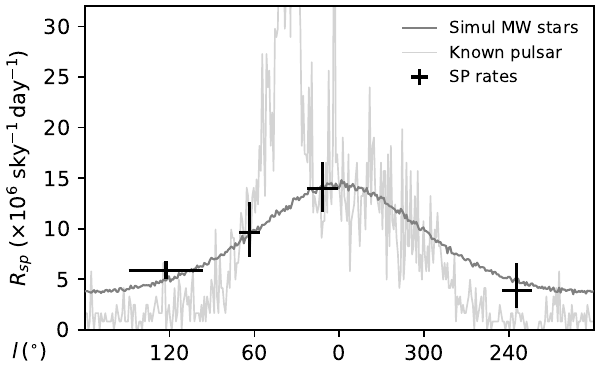}
   \caption{SP detection rate dependence on Galactic longitude. The longitude errors depict the longitude range of PTs with which the rates have been calculated. The rate errors are the 90\% Poisson errors as given in Table \ref{tab:sp_rates}. The simulated longitudinal distribution of stars in the Galactic disk is shown up to a distance of 2.2\,kpc from the Sun together with the longitudinal distribution of known pulsars from \psrcat. The stellar distribution is multiplied by a value $\rho_{0}$ (Eq. \ref{eq:stellar_dens}) such that it best describes the longitude rates. The pulsar distribution is given the same area as the stellar distribution.}
   \label{fig:sp_long_rates}
\end{figure}

To further test a possible Galactic origin of these candidates, the SP rates of the four longitude regions in Table \ref{tab:sp_rates} were plotted in Fig. \ref{fig:sp_long_rates}. The given errors are 90\% confidence limits assuming Poisson statistics. A longitude dependence of these rates is seen, with the rates clearly increasing towards the Galactic centre, indicating a Galactic connection.

In order to strengthen this observation, the distribution of stars in the Galactic disk\footnote{Only stars in the Galactic disk are considered, because the effects of the bulge are negligible at the stellar distances here under consideration.} was considered. Assuming the isolated SP candidates to have a stellar origin, their longitudinal detection rate relation should be describable with this stellar distribution. Therefore, the stellar density, $\rho$, at a projected distance, $R$, on the Galactic plane and from its centre, and height, $z$, from the Galactic plane was calculated as
\begin{equation} \label{eq:stellar_dens}
    \rho(R,z) = \rho_{0}\,e^{-R/H}\,\mathrm{sech}(z/h)^{2}\quad \mathrm{pc}^{-3}.
\end{equation}
In this equation, taken from \citet{nypv2001}, $H$ = 2.5\,kpc, $h$ = 200\,pc, and the age and mass dependence of $h$ is neglected. With Eq. \ref{eq:stellar_dens}, the stellar distribution of stars was simulated in the solar neighbourhood ($R_{\odot}$ = 8.5\,kpc, $z_{\odot}$ = 30\,pc). Therefore, a change in coordinate system was made to relate $R$ and $z$ with the distance, $d$, from the Sun:
\begin{align*}
    x &= R~\cos(\phi), \\
    y &= R~\sin(\phi), \\
    d &= \sqrt{x^{2} + (y - R_{\odot})^{2} + z^{2}},
\end{align*}
and $\phi$ is a randomly sampled angle between $\pi/2~\pm \arctan(d/R_{\odot})$. Simulated stars up to a distance $d$ and with a Galactic latitude, $b$, between $\pm15^{\circ}$ were binned by one longitudinal degree to obtain the distribution of local stars over Galactic longitude. The obtained distribution was then scaled to the rates plotted in Fig. \ref{fig:sp_long_rates} by taking a value for $\rho_{0}$ that minimises the two-dimensional distances between the rates and the curve of the stellar distribution. Doing this for various values of $d$, yielded a distance $d$ = 2.2\,kpc for which the simulated stellar distribution agrees best with the longitude rates. An arbitrary scaling was thus applied to the absolute scale of the stellar distribution to ease the comparison of its relative shape with that probed by the longitudinal SP detection rates.

Similar to the simulated stars, a histogram of all pulsars in \psrcat\ was made in bins of one degree of Galactic longitude. This histogram's area was given the same area as that of the simulated stars up to a distance of 2.2\,kpc, to enable these distributions' shapes to be compared. As can be seen, the shape of the Galactic pulsar population also agrees fairly well with the SP detection rate's longitudinal distribution. Caution must be taken, though, with the pulsar distribution, which contains many biases. For instance, the spike around $l \sim 30^{\circ}$ is caused by this part of the sky (the celestial equator) being visible for almost all radio observatories around the world. This part of the sky therefore harbours the most pulsar detections.

\begin{figure}
\centering
\includegraphics[width=0.48\textwidth]{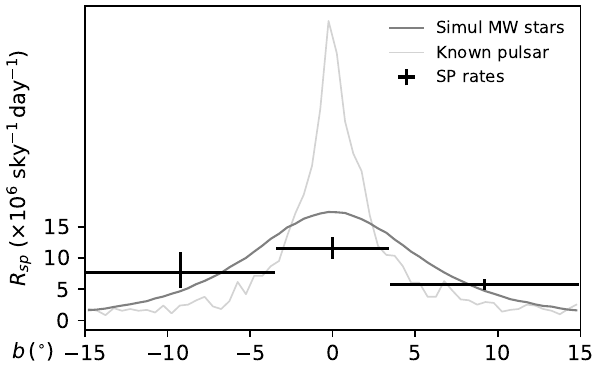}
   \caption{Similar to Fig. \ref{fig:sp_long_rates} but now depicting the SP detection rate dependence on Galactic latitude. The same simulated stellar distribution is shown as a function of Galactic latitude, together with the latitude distribution of known pulsars.}
   \label{fig:sp_lat_rates}
\end{figure}

If the isolated SP candidates indeed originate from a Galactic stellar population, their detection rates over Galactic latitude should agree with the simulated stellar distribution as well. These rates and distribution are plotted in Fig. \ref{fig:sp_lat_rates}, now considering the positive and negative parts of the mid-lat range as separate regions, as opposed to what was done in Table \ref{tab:sp_rates}. The same simulated distribution of stars up to a distance of 2.2\,kpc was binned by 0.5 latitude degrees, and scaled such that the total number of stars, the area underneath the curve, is the same for Figs. \ref{fig:sp_long_rates} and \ref{fig:sp_lat_rates}. The pulsar distribution was again added and given the same area as the fitted stellar distribution.

Over Galactic latitudes, these distributions describe the observed SP rates less well, though the stellar distribution comes close. This is especially true when considering that the same arbitrary scaling, multiplication by $\rho_{0}$ (Eq. \ref{eq:stellar_dens}), was applied to this distribution as the longitudinal stellar distribution, and that the rates may be composed of false positive candidates or extragalactic SPs. All isolated SPs from Table \ref{tab:sp_cands} were used to calculate the rates, even though some have a higher DM than the Galactic DM contribution estimated by the NE2001 model. This was done to not introduce potential biases through a DM selection of candidates and accommodate for possible errors in the electron density model. Re-calculating the rates with just those SPs with a DM < DM$_{gal}$, or a member count > 10, resulted in similar trends, but with too low statistical relevance to improve the reliability of the findings.

Nevertheless, the rates including all isolated SPs might show a flatter relation compared to the given distributions. This could indicate that the sources of isolated SPs are less bound to the Galactic plane and that the high longitude rate at the Galactic centre bleeds into the mid-lat region rates, causing them to attain a value closer to that of the low-lat rate. To disentangle such a possible effect, it would be best to calculate detection rates for separate longitude and latitude regions, but there are too few isolated SP candidates for this to result in statistically relevant rates. Alternatively, it might be more difficult to detect isolated SPs in the low-lat region, explaining the lower low-lat rate compared to the pulsar distribution. This is not unimaginable if the detection of these SPs is hampered by absorption and scattering in the Galactic plane. The current rates are not constraining enough to favour a stellar or pulsar model, but are suggestive of a faint Galactic SP population.

\subsection{SP discoveries} \label{sec:sp_disc}
HFS26 calculated the number of expected SP candidates above a given S/N threshold due to instrumental noise. Repeating this calculation for the entirety of the 5000 HTRU-North PTs processed yielded 0.3 expected candidates above an S/N of 8. Among the isolated SP candidates from HFS26 and those from Table \ref{tab:sp_cands} are, however, five SPs with an S/N > 8. With the recognition that for the HTRU-North data a member count above ten is indicative of true positive events, three of these (C109, C116, and C016) most likely represent authentic SPs since their member count is above 10. C109 was detected with a member count of 482. A re-examination of its frequency-time structure revealed it to be a SP of B0329+54. This pulse must have been very bright as the separation between the pulsar and C109's detection beam is 0.8 degrees, suggesting it was detected with the fourth sidelobe of its detection beam. The two other candidates are individually discussed below.

\subsubsection{RRAT J0404+53} \label{sec:C116}

\begin{figure}
\centering
\includegraphics[width=0.45\textwidth]{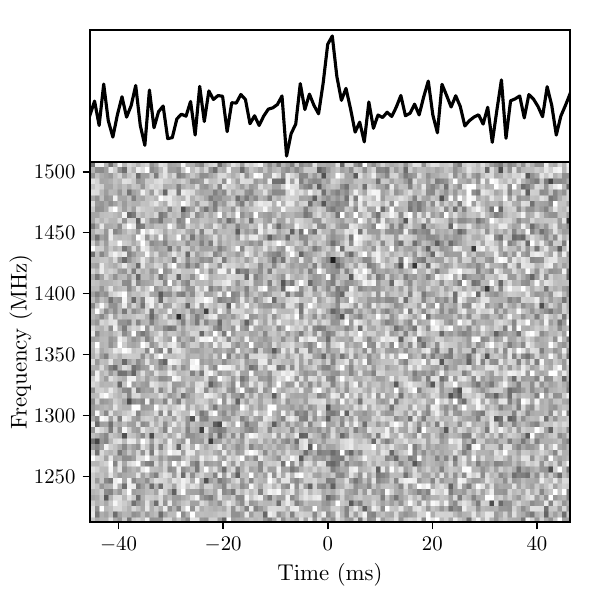}
   \vspace{-5pt}
   \caption{De-dispersed time series (top) and dynamic spectrum (bottom) of C116 originating from a new RRAT, J0404+53.}
   \label{fig:dyn_spec_c116}
\end{figure}

C116 was detected with a member count of 57 and a detection S/N of 8.0. According to the PSS, no known pulsar resides even within one degree of the Galactic position of its detection beam. Through iteratively de-dispersing its received emission with DM steps of 0.1\,\dm, an optimised S/N of 9.1 was obtained for this pulse with a DM of 73.0\,\dm\ and width of 2.5 $\pm$ 0.1\,ms. The increase in S/N is due to it being determined with a more optimal width and slightly improved DM, compared to \heim's S/N estimate. According to the NE2001 and YMW16 \citep{ymw2017} models, the total Galactic DM contribution in C116's line of sight is 198 and 292\,\dm, respectively. These estimates place the origin of this burst well within the Galaxy. Due to its S/N, high member count and convincing pulse profile (Fig. \ref{fig:dyn_spec_c116}), C116 is believed to originate from a new RRAT: J0404+53, at a position of RA = \ensuremath{{04}^\mathrm{h}{04}^\mathrm{m}} and Dec = +\ensuremath{{53}^\mathrm{\circ}{01}\mathrm{'}} with a circular error radius of 0.08$^\mathrm{\circ}$.

On October 6, 2025, this source was followed up with a gridding of 20 minutes per scan. Within the obtained data of these scans, no SPs of astrophysical origin were found. Although strongly affected by RFI, the non-detection places an upper limit on the source's burst rate of < 1.4\,h$^{-1}$. The data of these scans were also periodicity searched at C116's DM, but no average emission could be found. Nor could average emission be found in the HTRU-North PT in which this SP was detected. The low burst rate and the non-detection of a folded profile imply that this source is most likely a new RRAT.

J0404+53's Galactic latitude of 0.36$^\mathrm{\circ}$ places it almost exactly on the major axis of the Galactic plane, another observation that boosts confidence in this SP originating from a new Galactic source of pulsating radio emission. To truly confirm this to be the case, SPs from a similar position and with a similar DM need to be re-detected. With its low burst rate and S/N with Effelsberg, this requires a long exposure time with a sensitive radio telescope. Or, if J0404+53 is also visible between 800 and 400\,MHz, CHIME may detect it as its NE2001-estimated scattering time at 400\,MHz is just $\sim$0.08\,ms.

\subsubsection{FRB 20110220A} \label{sec:FRB}

\begin{figure}
\centering
\includegraphics[width=0.45\textwidth]{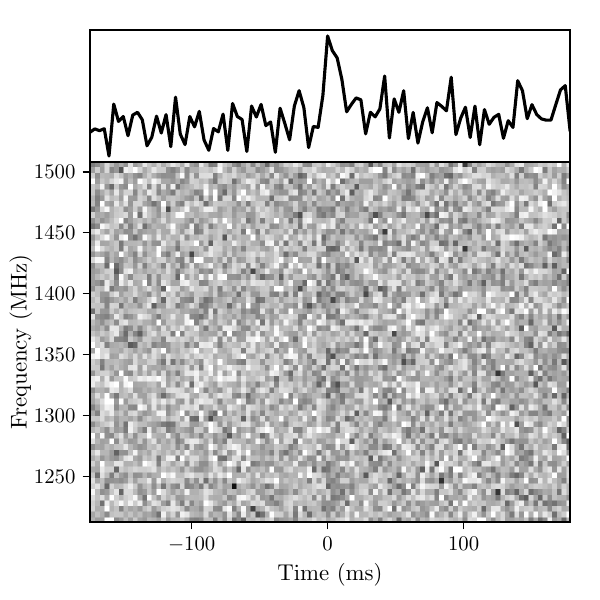}
   \vspace{-5pt}
   \caption{De-dispersed time series and dynamic spectrum of FRB\,20110220A.}
   \label{fig:dyn_spec_c016}
\end{figure}

C016 was initially detected with a member count of 28 and a detection S/N of 8.0. Through optimising its S/N, a DM of 501.0\,\dm\ and width of 11.9 $\pm$ 3.5\,ms was found for an optimised S/N of 9.3, which gives it a fluence of 0.6 $\pm$ 0.1\,Jy\,ms. With the current knowledge of the characteristics of the HTRU-North SP pipeline, especially the observation that a pulse member count > 10 most likely distinguishes real SPs from instrumental noise (Sect. \ref{sec:sp-trains}), this SP is considered a true positive pulse. The Galactic contribution to the DM at the position of C016 (RA = \ensuremath{{18}^\mathrm{h}{45}^\mathrm{m}} and Dec = +\ensuremath{{09}^\mathrm{\circ}{24}\mathrm{'}}) is 335 and 260\,\dm\ according to the NE2001 and YMW16 models, respectively. These DM estimates are backed by the highest DM of any known pulsar within a radius of two degrees around C016's position (using the PSS), that of J1851+0843g \citep{hzw+2025}, which has a DM of 246.3\,\dm. The DM excess is therefore 166 and 241\,\dm, likely placing C016 outside the Milky Way. Thus, C016, which was detected on 20-02-2011 at 05:33:23 UTC, is the first FRB discovered with Effelsberg. From now on, this SP is denoted as FRB\,20110220A.

Although its measured width is affected by intra-channel dispersion smearing of the order of $\sim$1\,ms, its inferred intrinsic width is still rather wide and does not provide a good indication about it being an FRB repeat burst or not \citep{pgk+2021}. In 15 minutes of follow-up data, obtained on August 19, it was not re-detected. No mention of an FRB detection within 2.5$^\mathrm{\circ}$ of the Galactic location of FRB\,20110220A could be found in either the Transient Name Server (TNS)\footnote{\url{https://www.wis-tns.org/}} or the CHIME/FRB first and second FRB catalogues~\citep{taa+2021,caa+2026}. This, together with its morphology as a simple Gaussian-shaped pulse (Fig. \ref{fig:dyn_spec_c016}), probably indicates FRB\,20110220A to resemble a one-off FRB.

To assess if an FRB detection was to be expected in the HTRU-North data, the expected number of FRBs was calculated using the all-sky rates from the Apertif and Parkes FRB samples. \citet{pvb+2025} found an FRB all-sky rate for Apertif, operating at 1.36\,GHz and above a fluence completeness threshold of 4.1\,Jy\,ms, of $R_{\mathrm{Apertif\_\textsc{frb}}}$ = 459$^{+208}_{-155}$\,sky$^{-1}$\,day$^{-1}$. This rate was first scaled to the fluence completeness threshold for morphology I bursts of the HTRU-North SP search obtained in HFS26 (\mbox{$F_{MI} \simeq$ 0.4\,Jy\,ms}) using the scaling relation given by \citet{jem+2019}:
\begin{equation} \label{eq:rate_scale}
    R(> F) = R_{0}\left(\frac{F}{F_{0}}\right)^{\alpha}\quad \mathrm{sky}^{-1}\,\mathrm{day}^{-1},
\end{equation}
where $R_{0}$ is the FRB all-sky rate at a fluence completeness threshold of $F_{0}$. Therefore, the above $R_{\mathrm{Apertif\_\textsc{frb}}}$ was taken for $R_{0}$ with $\alpha$ = -1.23 $\pm$ 0.06 as determined by \citet{pvb+2025}. The obtained result was then multiplied by the 0.02\,deg$^{2}$ field of view of a beam and the total observation time of the data processed. Excluding beam files strongly affected by RFI, 2309\,h of data remain. A number of 0.4$^{+0.2}_{-0.1}$ FRBs is expected to be present in the processed HTRU-North data according to the Apertif all-sky rate. This calculation was repeated with the all-sky FRB rate reported for Parkes \citep{bkb+2018}, $R_{\mathrm{Parkes\_\textsc{frb}}}$ = 1.7$^{+1.5}_{-0.9}$ $\times$ 10$^{3}$\,sky$^{-1}$\,day$^{-1}$, operating at 1.38\,GHz and above a fluence completeness threshold of 2\,Jy\,ms. Using the updated spectral index measured by \citet{jem+2019} of $\alpha$ = -1.18 $\pm$ 0.24 for the Parkes FRB sample, a number of 0.5$^{+0.5}_{-0.3}$ expected FRBs is found. These two numbers agree and show that detecting one FRB in the processed HTRU-North data is not unexpected.

\section{Discussion} \label{sec:discussion}
We have reported the discovery of a new FRB, FRB\,20110220A,  based on its DM excess compared to the Galactic DM estimates of the NE2001 and YMW16 models (Sect. \ref{sec:FRB}). However, these models contain significant uncertainties, as was pointed out by their creators and \citet{pfd2021}. The excesses are smaller than the estimated total Galactic DM contributions. It might therefore be possible that FRB\,20110220A actually has a Galactic origin whenever, for instance, H$\alpha$ or H$\beta$ regions increase the total electron column density in the line of sight towards the burst, just as for FRB\,010621 \citep{bm2014}. We consider FRB\,20110220A to be of extragalactic origin since the NE2001 and YMW16 models both place it outside the Galaxy, and no pulsar is known within a radius of two degrees around the FRB's position with a higher DM than the modelled Galactic DM estimates.

This discovery resulted from the analysis of one-fifth of the available HTRU-North data. Based on the more recent Apertif all-sky FRB rate of $R_{\mathrm{\textsc{frb}}}$ = 459$^{+208}_{-155}$\,sky$^{-1}$\,day$^{-1}$, Eq. \ref{eq:rate_scale}, and $\alpha$ = -1.23 $\pm$ 0.06 as determined by \citet{pvb+2025}, we predict there are 4.4$^{+2.0}_{-1.5}$ FRB discoveries waiting to be made. This is the number of FRBs estimated to be present in the remaining HTRU-North data (see Sect. \ref{sec:FRB}).

Alongside an FRB, Sect. \ref{sec:results} also presented the detection of several known pulsars and a new RRAT. The continued processing of the remaining HTRU-North data is also expected to find such new sources, like J2028+28 and J0404+53. An estimate on how many of these sources are expected to be found is, however, hard to give. Many variables affect the detection rate of these sources. A best estimate can nonetheless be obtained by extrapolating the two detections here reported to the remaining data available. Another 7.5$^{+5.6}_{-4.2}$ RRATs might then be discoverable with the completion of the HTRU-North SP search.

The above estimates likely represent a lower bound on the yield of the continued HTRU-North SP search. Although most SP candidates and SP trains were not re-detected in follow-up observations, this does not exclude them from having an astrophysical origin. These candidates were initially detected with S/N values near the applied S/N threshold under more favourable RFI conditions. Changes in the RFI environment have since reduced the effective bandwidth and usable observing time of Effelsberg’s 21 cm receiver, limiting its sensitivity to faint and infrequently emitting sources. Re-detections with other instruments or alternative Effelsberg receivers would therefore plausibly increase the number of confirmed detections from the HTRU-North SP search.

In light of the member counts of some of the candidates, including T142.0+9.46 (Table \ref{tab:sp_trains}) and T102.1+01.4 (Table E.2 from HFS26), more (re-)detections are expected. We have argued that the \heim\ member count is a good indication of a candidate's authenticity (Sect. \ref{sec:sp-trains}). For the HTRU-North data, the injection tests preformed by HFS26 showed that SP candidates with a member count > 10 are most often true positive events. It is good to point out that the indicativeness of \heim's member count might be different for other surveys, as its member count depends on a SP's DM and width, and how \heim\ is set to search for them.

From the observed isolated SP candidates, a faint Galactic SP population was recognised in Sect. \ref{sec:sp-cands}. Such a population would indicate that many faint SP-emitting sources exist within the Galaxy that have yet to be discovered. Their discovery will probably require more sensitive instruments, such as FAST and the Square Kilometre Array Observatory (SKAO; \citealt{dls+2022}), though, this tentative prognosis can be made more robust when a similar Galactic longitude and latitude relation in the detection rates of isolated SP candidates is found by other SP searching instruments. Until confirmed by another survey, the here presented Galactic population remains tentative.

If confirmed, at least a part of the candidates, from which the population is derived, are shown to be real. The question then arises what origin these SPs might have. In line with Occam's razor, the observed isolated SP candidates most likely have a neutron star origin. However, this might have implications for the total neutron star population and present a birthrate problem as discussed by \citet{kk2008}. Since the detection rates of Sect. \ref{sec:sp-cands} can also be described with a distribution of local stars in the Galaxy, it might be possible that they originate from flaring stars \citep{bbb+2001,ber2002,lsm2014}, for instance. The current rates do not favour either of the two proposed progenitor models. Confirmation of the existence of the population and future discoveries of underlying faint SPs are therefore needed to determine their true origin.

\newpage
\section{Conclusions} \label{sec:conclusions}
We have presented a SP search of 5000 PTs of the HTRU-North survey. Insights gained from this search and follow-up observations performed with the 21\,cm receiver of the 100\,m Effelsberg Radio Telescope yielded the following discoveries:
\begin{itemize}[leftmargin=0pt]
    \item FRB\,20110220A: Effelsberg's\;first FRB discovery (Sect. \ref{sec:FRB});
    \item J0404+53: a new RRAT (Sect.~\ref{sec:C116});
    \item J2028+28: the first L-band detection of this RRAT (Sect.~\ref{sec:re-detections});
    \item Eight new candidate SP trains (Table~\ref{tab:sp_trains});
    \item 272 isolated SP candidates (Table~\ref{tab:sp_cands}).
\end{itemize}
We determined all-sky detection rates from the isolated SP candidates that show a Galactic latitude and longitude dependence (Sect. \ref{sec:sp-cands}). These dependences suggest the existence of a faint Galactic SP population. Other research groups are encouraged to reproduce and confirm these findings.

Also, 35 known pulsars and 3 known RRATs (including J2028+28) were re-discovered in the newly analysed PTs. About $\sim$21\% of the available HTRU-North data have now been searched for SPs. With a relatively low degree of RFI contamination, it is promising to continue searching these data for SPs. This is especially the case since the RFI environment at the Effelsberg site has evolved strongly over time, which probably caused only a small fraction of followed-up candidates to be re-detected. In the HTRU-North data, many new sources remain hidden that might be discoverable with a level of ease that has become hard to match, motivating the continued analysis of these data.

\section*{Data availability}
Full Table \ref{tab:sp_cands} is available at the CDS via \url{https://cdsarc.cds.unistra.fr/viz-bin/cat/J/A+A/708/A199}.

\begin{acknowledgements}
This work was supported by the Netherlands Organization for Scientific Research (NWO) Spinoza Prize awarded to Heino Falcke, and is based on observations with the 100-m telescope of the MPIfR (Max-Planck-Institut f\"ur Radioastronomie) at Effelsberg. Part of the observations were performed with the new EDD system developed and maintained by the electronics division at the MPIfR. The authors wish to thank this MPIfR team and all the observers who helped accumulate the HTRU-North data over the past years, and so enabled the research presented here. We also acknowledge use of the CHIME/FRB Public Database, provided at \url{https://www.chime-frb.ca/} by the CHIME/FRB Collaboration.
\end{acknowledgements}

\vfill

\bibliographystyle{aa}
\bibliography{master}

\begin{appendix}

\twocolumn[\section{Additional figures and tables} \label{sec:A.figs&tabs}]

\begin{figure}[h!]
    \hspace{0.8cm}
    \includegraphics[height=0.66\linewidth, angle=-90]{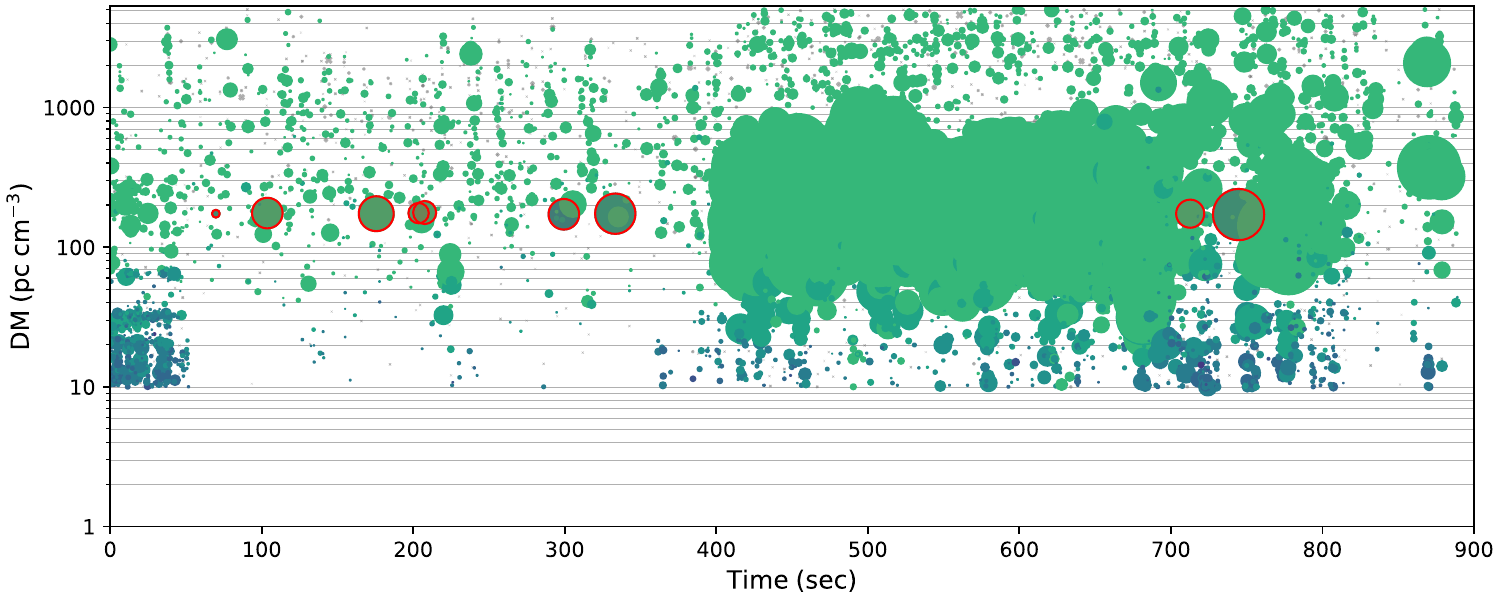}
    \caption{DM-time plot showing all the candidates for scan 2 of T054.9-03.0 found by \heim. The detected SPs from B1942+17 are highlighted in red. These pulses were detected during the moments of reduced RFI contamination between $\sim$100 and 400\,s and around $\sim$750\,s.}
    \label{fig:det_plot_B1942+17}
\end{figure}

\vfill

\begin{table}[h!]
  \small
  \caption{Isolated SP candidates in the production PTs, sorted by the Galactic longitude (GL) and latitude (GB) of the candidates.}
  \label{tab:sp_cands}
  \centering
  \begin{tabular}{l c c c c c r c c r l}
    \hline
    \noalign{\smallskip}
    Name&GL&GB&RA&Dec&S/N&DM&DM$_{Gal}$&Width&Mem.&Note\\
    & (deg) & (deg) & (deg) & (deg) && \multicolumn{2}{c}{(\dm)} & (ms) & &\\
    \noalign{\smallskip}
    \hline
    \noalign{\smallskip}
    C141 & 6.25 & 6.86 & 263.62 & -20.02 & 6.87 & 58.4  & 359.8 & 1.7 & 10 &\\
    C142 & 6.50 & 8.59 & 262.21 & -18.88 & 7.19 & 580.4 & 283.8 & 3.5 & 8  &\\
    C143 & 6.75 & 8.45 & 262.48 & -18.75 & 7.21 & 247.4 & 288.8 & 7.0 & 6  &\\
    \multicolumn{11}{c}{\ldots} \\
    \noalign{\smallskip}
    \hline
  \end{tabular}
  \tablefoot{The full table of 272 candidates is available at the CDS.\\The last column indicates whether a detected candidate may originate from a known source.}
\end{table}

\begin{table}[h!]
\small
\caption{Revised SP train detections in the analysed HTRU-North analysis and production PTs, sorted by perceived genuineness.}
\label{tab:sp_trains}
\centering
\begin{tabu}{l c c c l c r}
    \hline
    \noalign{\smallskip}
    & GL & GB & S/N & DM & Width & Mem.\\
    & (deg) & (deg) & & (\dm) & \multicolumn{1}{c}{(ms)} & \\
    \noalign{\smallskip}
    \hline
    \noalign{\smallskip}
    \rowfont{\bfseries}
    \multirow{4}{*}{\rotatebox[origin=c]{90}{{\tiny T142.0+9.46}}} & 142.0 & 9.46 & & \multicolumn{3}{l}{57.6} \\
    \rowfont{\tiny}
    & \multicolumn{2}{r}{\textsc{i}.} & 7.34 & 57.2 & 1.7 & 17 \\
    \rowfont{\tiny}
    & \multicolumn{2}{r}{\textsc{ii}.} & 7.54 & 58.0 & 0.9 & 20 \\
    & & & & & & \\
    \noalign{\medskip}
    \hline
    \noalign{\smallskip}
    \rowfont{\bfseries}
    \multirow{4}{*}{\rotatebox[origin=c]{90}{{\tiny T125.6+0.00}}} & 125.63 & 0.00 & & \multicolumn{3}{l}{35.0} \\
    \rowfont{\tiny}
    & \multicolumn{2}{r}{\textsc{i}.} & 7.45 & 37.5 & 0.9 & 31 \\
    \rowfont{\tiny}
    & \multicolumn{2}{r}{\textsc{ii}.} & 6.85 & 32.5 & 0.4 & 6 \\
    & & & & & & \\
    \noalign{\medskip}
    \hline
    \noalign{\smallskip}
    \rowfont{\bfseries}
    \multirow{4}{*}{\rotatebox[origin=c]{90}{{\tiny T123.8-0.79}}} & 123.75 & -0.79 & & \multicolumn{3}{l}{31.3} \\
    \rowfont{\tiny}
    & \multicolumn{2}{r}{\textsc{i}.} & 6.96 & 35.0 & 3.5 & 20 \\
    \rowfont{\tiny}
    & \multicolumn{2}{r}{\textsc{ii}.} & 6.76 & 27.5 & 0.9 & 5 \\
    & & & & & & \\
    \noalign{\medskip}
    \hline
    \noalign{\smallskip}
    \rowfont{\bfseries}
    \multirow{4}{*}{\rotatebox[origin=c]{90}{{\tiny T127.1+9.53}}} & 127.12 & 9.53 & & \multicolumn{3}{l}{67.5} \\
    \rowfont{\tiny}
    & \multicolumn{2}{r}{\textsc{i}.} & 7.02 & 69.2 & 3.5 & 12 \\
    \rowfont{\tiny}
    & \multicolumn{2}{r}{\textsc{ii}.} & 6.98 & 65.7 & 0.9 & 6 \\
    & & & & & & \\
    \noalign{\medskip}
    \hline
    \noalign{\smallskip}
    \rowfont{\bfseries}
    \multirow{4}{*}{\rotatebox[origin=c]{90}{{\tiny T137.1+9.82}}} & 137.13 & 9.82 & & \multicolumn{3}{l}{47.0} \\
    \rowfont{\tiny}
    & \multicolumn{2}{r}{\textsc{i}.} & 6.83 & 47.1 & 1.7 & 6 \\
    \rowfont{\tiny}
    & \multicolumn{2}{r}{\textsc{ii}.} & 6.95 & 46.9 & 1.7 & 8 \\
    & & & & & & \\
    \noalign{\medskip}
    \hline
    \noalign{\smallskip}
    \rowfont{\bfseries}
    \multirow{4}{*}{\rotatebox[origin=c]{90}{{\tiny T161.1+6.06}}} & $^{*}$161.13 & 6.06 & & \multicolumn{3}{l}{58.6} \\
    \rowfont{\tiny}
    & \multicolumn{2}{r}{\textsc{i}.} & 7.64 & 57.1 & 0.4 & 13 \\
    \rowfont{\tiny}
    & \multicolumn{2}{r}{\textsc{ii}.} & 7.38 & 60.1 & 0.9 & 5 \\
    & & & & & & \\
    \noalign{\medskip}
    \hline
    \noalign{\smallskip}
    \rowfont{\bfseries}
    \multirow{4}{*}{\rotatebox[origin=c]{90}{{\tiny T106.9+6.82}}} & 106.94 & 6.82 & & \multicolumn{3}{l}{42.6 $\pm$ 4.4 \quad P - 0.080\,(s)} \\
    \rowfont{\tiny}
    & \multicolumn{2}{r}{\textsuperscript{\textdagger}\textsc{i}.} & 6.84 & 42.9 & 0.9 & 3 \\
    \rowfont{\tiny}
    & \multicolumn{2}{r}{\textsc{ii}.} & 6.80 & 46.8 & 1.7 & 5 \\
    \rowfont{\tiny}
    & \multicolumn{2}{r}{\textsc{iii}.} & 7.50 & 38.0 & 0.2 & 7 \\
    \noalign{\medskip}
    \hline
    \noalign{\smallskip}
    \rowfont{\bfseries}
    \multirow{4}{*}{\rotatebox[origin=c]{90}{{\tiny T112.9+2.71}}} & 112.9 & 2.71 & & \multicolumn{3}{l}{65.9 $\pm$ 7.9 \quad P - 0.086\,(s)} \\
    \rowfont{\tiny}
    & \multicolumn{2}{r}{\textsc{i}.} & 6.73 & 74.3 & 1.7 & 4 \\
    \rowfont{\tiny}
    & \multicolumn{2}{r}{\textsc{ii}.} & 6.79 & 64.6 & 0.2 & 3 \\
    \rowfont{\tiny}
    & \multicolumn{2}{r}{\textsuperscript{\textdagger}\textsc{iii}.} & 6.80 & 58.7 & 0.9 & 4 \\
    \noalign{\medskip}
    \hline
\end{tabu}
\tablefoot{
\begin{itemize}
    \item[$^{\dagger}$] Detected in an adjacent beam.
    \item[$^{*}$] Under the revised definition, a SP train from the analysis PTs.
\end{itemize}
}
\end{table}

\end{appendix}

\end{document}